\documentclass[sn-mathphys,Numbered]{sn-jnl}

\usepackage{graphicx}%
\usepackage{multirow}%
\usepackage{amsmath,amssymb,amsfonts}%
\usepackage{amsthm}%
\usepackage{mathrsfs}%
\usepackage[title]{appendix}%
\usepackage{xcolor}%
\usepackage{textcomp}%
\usepackage{manyfoot}%
\usepackage{booktabs}%
\usepackage{algorithm}%
\usepackage{algorithmicx}%
\usepackage{algpseudocode}%
\usepackage{listings}%

\theoremstyle{thmstyleone}%
\theoremstyle{thmstyletwo}%

\theoremstyle{thmstylethree}%

\begin{document}

\title[Article Title]{Navigate Biopsy with Ultrasound under Augmented Reality Device: Towards Higher System Performance}


\author[1]{\fnm{Haowei} \sur{Li}}\email{lhw22@mails.tsinghua.edu.cn}
\author[2]{\fnm{Wenqing} \sur{Yan}}\email{ywq22@mails.tsinghua.edu.cn}
\author[1]{\fnm{Jiasheng} \sur{Zhao}}\email{zjs20@mails.tsinghua.edu.cn}
\author[2]{\fnm{Yuqi} \sur{Ji}}\email{jyq22@mails.tsinghua.edu.cn}
\author[3]{\fnm{Long} \sur{Qian}}\email{long@medivis.com}
\author[1]{\fnm{Hui} \sur{Ding}}\email{dinghui@tsinghua.edu.cn}
\author[4,5]{\fnm{Zhe} \sur{Zhao}}\email{zhaozhao\_02@163.com}
\author*[1]{\fnm{Guangzhi} \sur{Wang}}\email{wgz-dea@tsinghua.edu.cn}


\affil*[1]{\orgdiv{Biomedical Engineering}, \orgname{Tsinghua University}, \orgaddress{\street{Shuang Qing Road}, \city{Beijing}, \postcode{100084}, \state{Beijing}, \country{China}}}
\affil*[2]{\orgdiv{School of Medicine}, \orgname{Tsinghua University}, \orgaddress{\street{Shuang Qing Road}, \city{Beijing}, \postcode{100084}, \state{Beijing}, \country{China}}}
\affil*[3]{\orgname{Medivis Inc.}, \orgaddress{\street{920 Broadway}, \city{New York}, \postcode{10010}, \state{New York}, \country{USA}}}
\affil*[4]{\orgdiv{School of Clinical Medicine}, \orgname{Tsinghua University}, \orgaddress{\street{Shuang Qing Road}, \city{Beijing}, \postcode{100084}, \state{Beijing}, \country{China}}}
\affil*[5]{\orgdiv{Orthopedics}, \orgname{Beijing Tsinghua Chang Gung Hospital}, \orgaddress{\street{Li Tang Road}, \city{Beijing}, \postcode{100043}, \state{Beijing}, \country{China}}}





\abstract{
\textbf{Purpose:} 
Biopsies play a crucial role in determining the classification and staging of tumors. Ultrasound is frequently used in this procedure to provide real-time anatomical information. Using augmented reality (AR), surgeons can visualize ultrasound data and spatial navigation information seamlessly integrated with real tissues. This innovation facilitates faster and more precise biopsy operations.

\textbf{Methods:} 
We developed an AR biopsy navigation system with low display latency and high accuracy. Ultrasound data is initially read by an image capture card and streamed to Unity via net communication. In Unity, navigation information is rendered and transmitted to the HoloLens 2 device using holographic remoting. Retro-reflective tool tracking is implemented on the HoloLens 2, enabling simultaneous tracking of the ultrasound probe and biopsy needle. Distinct navigation information is provided during in-plane and out-of-plane punctuation. To evaluate the effectiveness of our system, we conducted a study involving ten participants, for puncture accuracy and biopsy time, comparing to traditional methods.

\textbf{Results:} 
Our proposed framework enables ultrasound visualization in AR with only $16.22\pm11.45ms$ additional latency. Navigation accuracy reached $1.23\pm 0.68mm$ in the image plane and $0.95\pm 0.70mm$ outside the image plane. Remarkably, the utilization of our system led to $98\%$ and $95\%$ success rate in out-of-plane and in-plane biopsy.

\textbf{Conclusion:} 
To sum up, this paper introduces an AR-based ultrasound biopsy navigation system characterized by high navigation accuracy and minimal latency. The system provides distinct visualization contents during in-plane and out-of-plane operations according to their different characteristics. Use case study in this paper proved that our system can help young surgeons perform biopsy faster and more accurately.
}




\maketitle

\section{Introduction} \label{sec1}
Tumor biopsy is an essential clinical procedure for accurately classifying and staging tumors with minimal invasiveness.
Intraoperative surgical navigation is necessary during this procedure to ensure precise puncturing of the target area while avoiding damage to critical organs.
Previous research has explored the use of computed tomography (CT) \cite{baratella2022accuracy, kim2022diagnostic} and ultrasound \cite{walter2022ultrasound, schonewille2022tracked} for this purpose.
Ultrasound navigation, in particular, offers real-time imaging without radiation exposure.
Surgeons can freely adjust the placement of the ultrasound probe dynamically to visualize organs and tissues from various angles and depths.
Consequently, it is widely adopted in biopsy of liver cancer \cite{huang1996ultrasound}, breast cancer \cite{houssami2011preoperative}, bone cancer \cite{chira2017ultrasound}, etc.

However, ultrasound navigation has limitations, mainly generated by the restricted imaging area and limited information outside the ultrasound image.
In ultrasound-guided biopsy, surgeons punctuate in-plane or out-of-plane \cite{meiser2016comparison}.
In in-plane navigation, the biopsy needle is aligned with the image plane, necessitating precise adjustments of the ultrasound probe and biopsy needle to maintain the trajectory within the image plane.
Out-of-plane navigation offers more flexibility, but presents challenges in tracking the needle's tip as it moves, making it difficult to pinpoint the target accurately.
These limitations result in repeated punctures, extended procedure times, decreased biopsy accuracy, and inadequate tissue sampling \cite{choi2011factors}.

To address these limitations and enhance spatial information in ultrasound-guided biopsies, various methods have been proposed.
England et al. introduced a mechanical structure to constrain the biopsy needle to the image plane \cite{england2019value}.
Therefore, spatial relationship between biopsy needle and ultrasound image can be ensured.
Other approaches incorporate optical and electromagnetic tracking systems to track ultrasound probe and biopsy needle in three dimensional space \cite{czajkowska2018biopsy,hakime2012electromagnetic}.
The ultrasound image and biopsy needle can then be displayed in three-dimensional space for spatial information.
Surgical robots have also been integrated to assist in spatial information provision \cite{MAHMOUD201889}.
However, these solutions complicate the already crowded operating room and introduce potential occlusion issues.
On the other hand, the navigation information is displayed on external 2D screens, which lacks three-dimensional spatial perception and leads to hand-eye discoordination.

Augmented reality (AR) emerges as a promising technology that combines three-dimensional visualization with spatial information.
Such technology enables building integrated systems with optical sensors for target tracking and virtual display aligned with the real world \cite{rabbi2013survey}.
Consequently, AR has been applied to surgical navigation in diverse contexts, including neurosurgery \cite{meola2017augmented}, orthopedics \cite{laverdiere2019augmented}, and dental procedures \cite{kwon2018augmented}.
In this study, we employ augmented reality to enhance ultrasound biopsy navigation, providing improved spatial information, intuitive visualization of navigation data, and ultimately, improving biopsy accuracy, reducing procedure time, and enhancing safety in tumor biopsies.

\section{Related Work}
\subsection{Visualization of Ultrasound in Augmented Reality}
Visualization of ultrasound images in augmented reality has been discussed for a long time.
In the late 1990s, video-see-through headsets were employed for ultrasound visualization \cite{sauer2001augmented, state1996technologies}.
These setups typically involved connecting infrared tracking cameras, stereo RGB cameras, and ultrasound scanners to separate image capture cards on a single computer for tracking, processing, and rendering.
A notable drawback of this approach was the need for multiple cables and complicated setup.
Another approach used an additional screen attached to the ultrasound probe and a half-silvered mirror to enable in-situ ultrasound visualization without the need for a headset \cite{wu2005psychophysical}. 
However, this method increased the size of the probe and could only display contents on a specific plane.

\subsection{Recent Studies}
In recent years, the advent of commercial AR headsets with integrated tracking, data processing, and display capabilities has significantly improved the field. 
These headsets eliminate the need for cables, as they can process data directly on the device, making them more suitable for clinical scenarios.
However, the high level of integration comes with resource limitations, posing challenges to constructing AR ultrasound navigation systems with both high tracking accuracy and low-latency display.

Kuzhagaliyev et al. employed an external optical tracking device to simultaneously track the AR headset, needle electrodes, and ultrasound probe to obtain spatial information \cite{kuzhagaliyev2018augmented}.  
Ultrasound images were streamed to the headset for rendering and display.
While this approach improved biopsy accuracy from $7.4mm$ to $4.9mm$ \cite{ruger2020ultrasound}, it introduced additional calibration steps and potential errors due to the integration of the tracking system.
Trong et al. used the RGB sensor on AR headsets to track image targets fixed to the ultrasound probe \cite{nguyen2022holous}. 
Ultrasound images were streamed to the headset as PNG files.
This method eliminated the need for an external tracking device but introduced an additional latency of $80ms$ and limited the frame rate to 25 frames per second due to on-device resource constraints.
Von et al. improved tracking accuracy by using infrared tool tracking with HoloLens 2's depth camera \cite{von2022ultrarsound}.
This approach also offered performance benefits due to lower resource consumption during tracking. 
Costa conducted an evaluation of on-device resource usage for ultrasound visualization, revealing that data streaming took an average of $23.681\pm0.637ms$, while rendering took $14.247\pm0.653ms$ \cite{costa2023augmented}.
In contrast, remote rendering \cite{microsoft_holographic_remoting}, a method that involves rendering on a high-performance computer and streaming the compressed output to the AR device, can achieved extremely low latency by saving time on data streaming and rendering.
Previous work have adopted remote rendering for three-dimensional ultrasound visualization \cite{von2023augmenting}, but highlighted the challenges in implementing tracking methods with remote rendering.

\subsection{Our Contributions}
Compared to prior research, our main contribution can be summarized as follows.
\begin{itemize}
    \item We have constructed a system structure employing remote rendering to minimize display latency while preserving accurate target tracking capabilities.
    \item We have developed an algorithm to simultaneous track the ultrasound probe and the biopsy needle, even when part of the tool is occluded.
    \item We have proposed distinct visualization methods for in-plane and out-of-plane punctuation and validated the effectiveness of AR-assisted ultrasound biopsy navigation through use case studies.
\end{itemize}

\section{Methods} \label{sec2}

\begin{figure}[t]
    \centering
        \includegraphics[width=0.85\columnwidth]{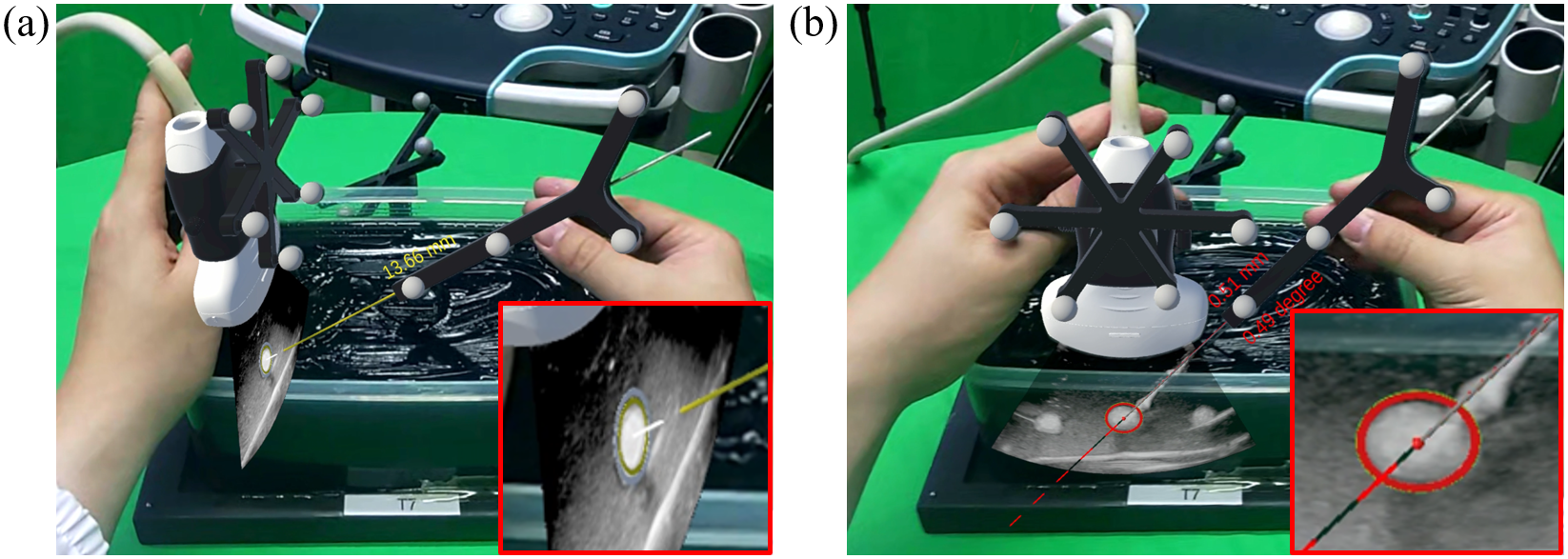}
    \caption{Biopsy navigation using ultrasound information overlayed on the real world. (a) Out-of-plane navigation. (b) In-plane navigation. Different visual contents are provided in these two modes to improve biopsy accuracy.}
    \label{fig:VirtualAlignedRealScene}
\end{figure}

\subsection{System Structure}
\begin{figure}[htpb]
    \centering
        \includegraphics[width=0.7\columnwidth]{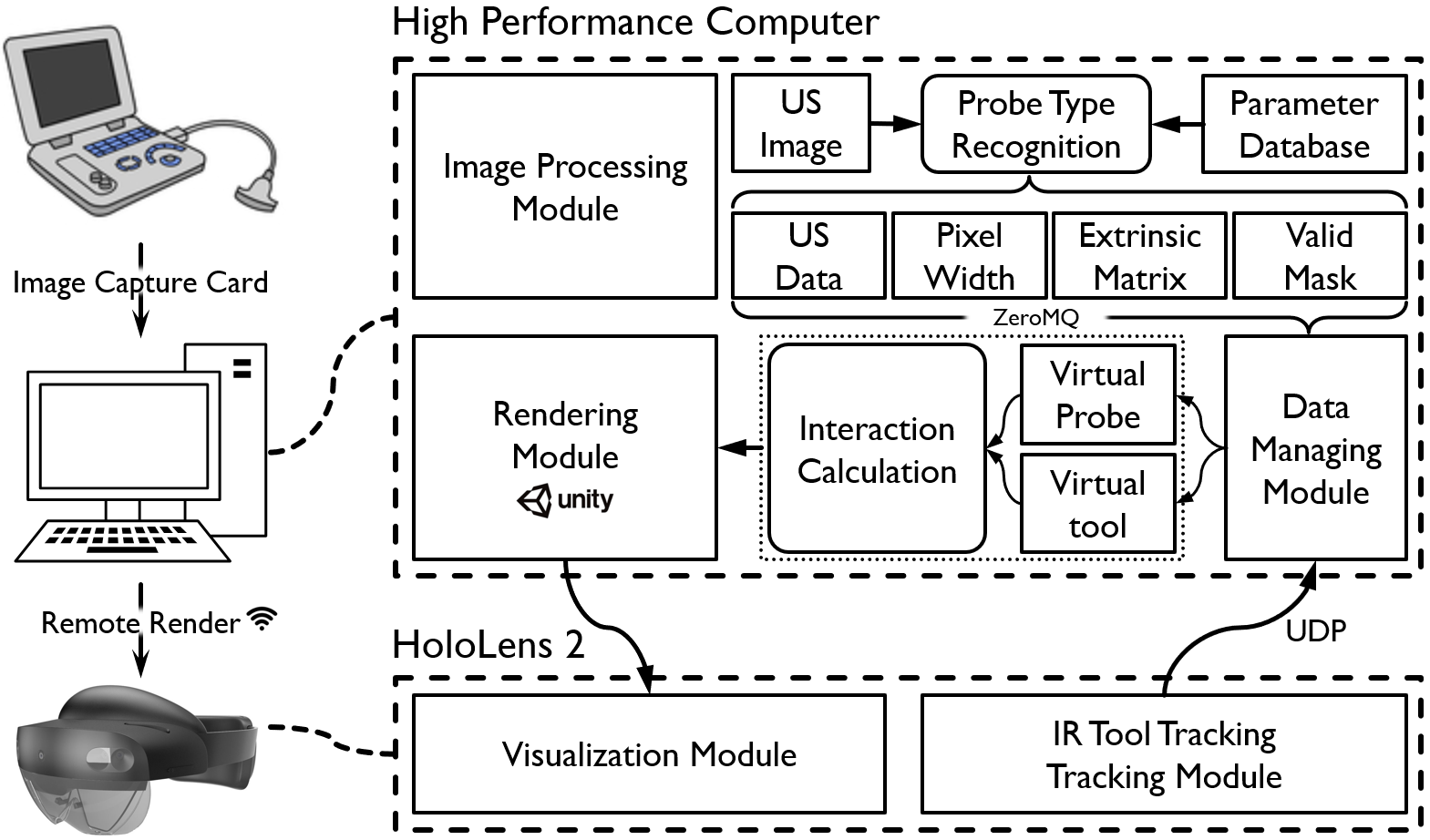}
    \caption{Structure of proposed ultrasound biopsy navigation system. Ultrasound images are captured and processed by a high-performance computer. Net communication synchronizes ultrasound data to Unity, which is rendered and streamed to HoloLens 2 through holographic remoting. Infrared tool tracking is implemented on the AR headset.}
    \label{fig:SystemStructure}
\end{figure}

The architecture of our proposed AR ultrasound navigation system is illustrated in Fig. \ref{fig:SystemStructure}. 
We employ the Resona 6 ultrasound system from Mindray (ShenZhen, China) for high-quality ultrasound images. 
To ensure a wide field of view during procedures, we utilize a convex probe (SC5-1U) with imaging depths ranging from $50mm$ to $400mm$.
As the ultrasound system does not support data acquisition through the network, we integrate an image capture card for ultrasound data transmission.

To minimize latency and ensure stable ultrasound image streaming, we employ a two-step approach.
Firstly, the ultrasound images are sent to a Unity project for rendering.
Subsequently, the rendered results are streamed to the HoloLens 2 device using holographic remoting \cite{microsoft_holographic_remoting}.
In this way, the streaming and rendering of ultrasound images are maintained within a single high-performance computer (HPC), where bandwidth, latency, and stability can be guaranteed.
Wireless connection between headset and the computer is only used to synchronize render results, where redundant information is compressed during rendering.

The infrared tool tracking module is implemented directly on the AR headsets using DirectX with C++.
Therefore, sensor data is directly accessed by the algorithm, eliminating additional API or data streaming overhead.
We employ the User Datagram Protocol (UDP) for low-latency streaming of tracking results to the HPC.
To further enhance performance, data reception is executed asynchronously through a custom C++ plugin, integrated into Unity as a dynamic link library (.DLL).

\subsection{Infrared Tool Tracking}
Infrared tool tracking using the HoloLens depth camera follows the principles outlined in a previous study \cite{von2022ultrarsound}.
The high reflectivity of retro-reflective spheres under the Articulated HAnd Tracking (AHAT) camera enables the fast localization of each marker, while pre-defined unique tool shape and distance information is used to identify each tool.
In this paper, our contribution mainly lies in 1. Simultaneously track ultrasound probe and biopsy needle, 2. Development of an algorithm that enables tool detection even when part of each tool is occluded.

Two different infrared tools are fixed on the ultrasound probe and biopsy needle, where the distance information between the marker balls can be extracted from the design. 
Potential matches between marker groups $(M_i, M_j) \to (S_m, S_n)$ are identified based on the difference in distances being within a specified range $||M_iM_j| - |S_mS_n|| < d$.
For each infrared tool, a Depth First Search (DFS) algorithm is employed to search for matches, utilizing previous matching information to inform the search for the next possible target.
To address the challenge of marker occlusion and maintain stable tool tracking, we set a maximum occlusion number for each tool.
During the search, whenever current marker is possibly occluded, it is flagged for matching of the next marker, creating an additional searching branch.
To ensure efficient searching, we employ a maximum matching rule, terminating the DFS when no possible match with less occlusion is found.
The conflicts between different potential marker matching results, generated by the ambiguity of side-length information, are then solved following the maximum matching minimum error rule to generate final searching results for pose estimation.

\subsection{Ultrasound Probe Parameter Identification}
\begin{figure}[htpb]
    \centering
        \includegraphics[width=0.65\columnwidth]{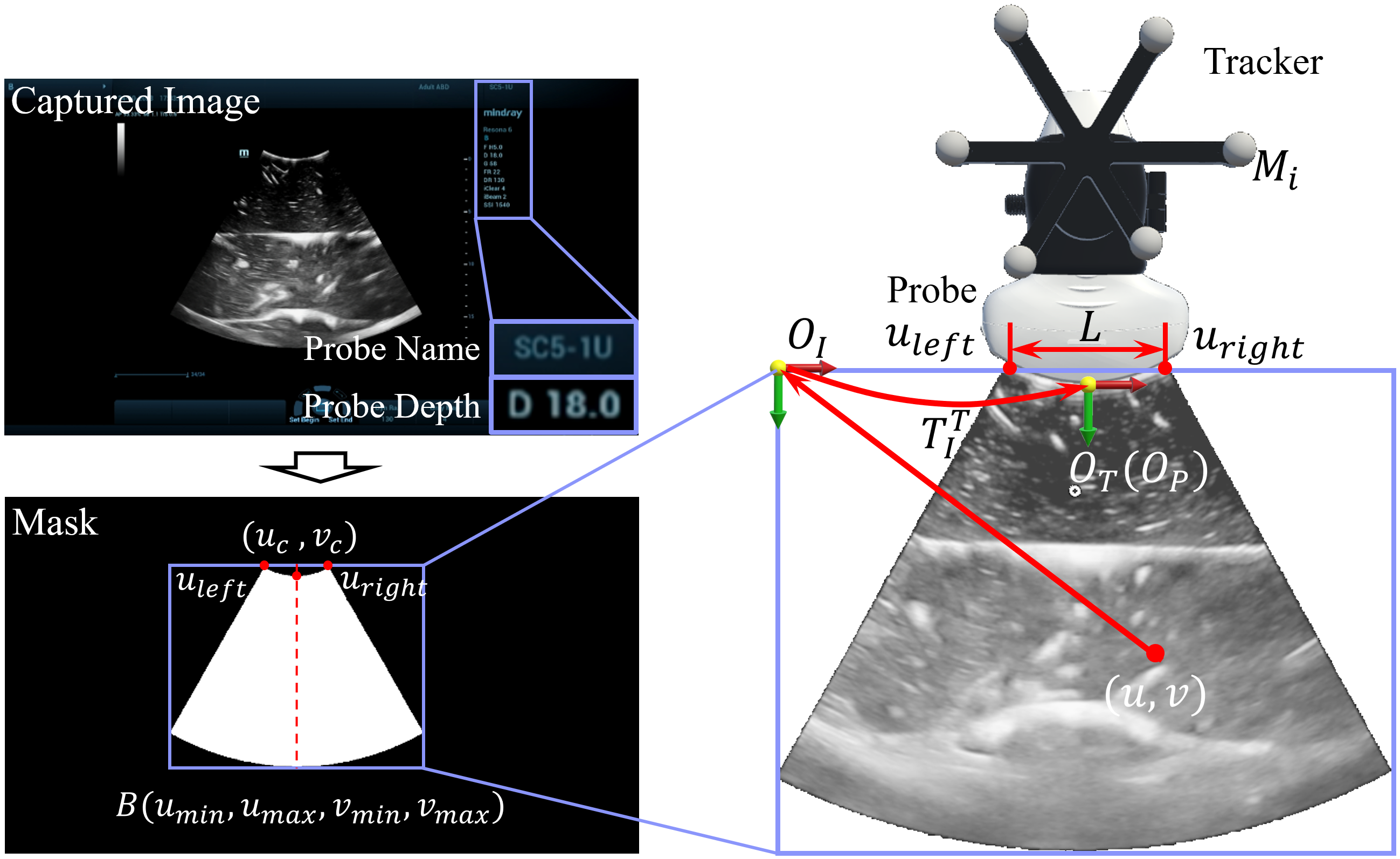}
    \caption{Registration of ultrasound image and infrared tool. The spaces of the ultrasound probe and infrared tool are aligned during designing. Transform from the ultrasound image to the probe is calculated using the shape of the valid imaging area.}
    \label{fig:UltrasounCalibration}
\end{figure}

Accurate spatial positioning of a given pixel $(u,v)$ from an ultrasound image with respect to the infrared tool is essential during AR visualization.
The alignment of the spaces occupied by the infrared tool ($O_T$) and the ultrasound probe ($O_P$) is maintained through tool design (see Fig.\ref{fig:UltrasounCalibration}). 
Restricted by probe property, the ultrasound image is generated in the $x-y$ plane, while there is no rotation between the image and the probe.
To analyze the pixel width $pw$ and relative shift $(v_x,v_y)$ between these two spaces, the binary mask of valid data area is first extracted, together with minimum bounding area $B(u_{min},u_{max},v_{min},v_{max})$.
The corners at the top of the mask are then identified for x position $u_{left},u_{right}$.
Providing the width of the sensors in the ultrasound $L$, the pixel width can be estimated as $pw=L/(u_{right}-u_{left})$.
The origin point of the probe on the image ($u_c, v_c$) is calculated as the first valid pixel in the middle.
Therefore, the extrinsic transform from the image to the infrared tool can be expressed as:
\begin{equation}
    T_I^T=
    \begin{bmatrix}
        pw & 0 & 0 & -pw(u_c-u_{min}) \\
        0 & pw & 0 & -pw(v_c-v_{min}) \\
        0 & 0 & 1 & 0 \\
        0 & 0 & 0 & 1
    \end{bmatrix}
\end{equation}

During AR visualization, the probe parameter is identified using the text located on the side of the ultrasound image.
Ultrasound data within the minimum effective bounding area ($B$) are streamed to Unity along with the mask.
Invalid pixels are set to transparent during rendering, while a large depth value is written to ensure accurate collision calculation.

Notably, this method works for both convex and linear probes, with the only change being the substitution of the corner points $(u_{left}, u_{right})$, used for pixel width calculation, with the bounding box borders $(u_{min}, u_{max})$.

\subsection{Navigation Visualization}
Two common methods in ultrasound guided biopsy are In-plane and out-of-plane punctuation.
Distinct visualization methods are designed to provide different guidance cues.


\begin{figure}[htpb]
    \centering
        \includegraphics[width=1.0\columnwidth]{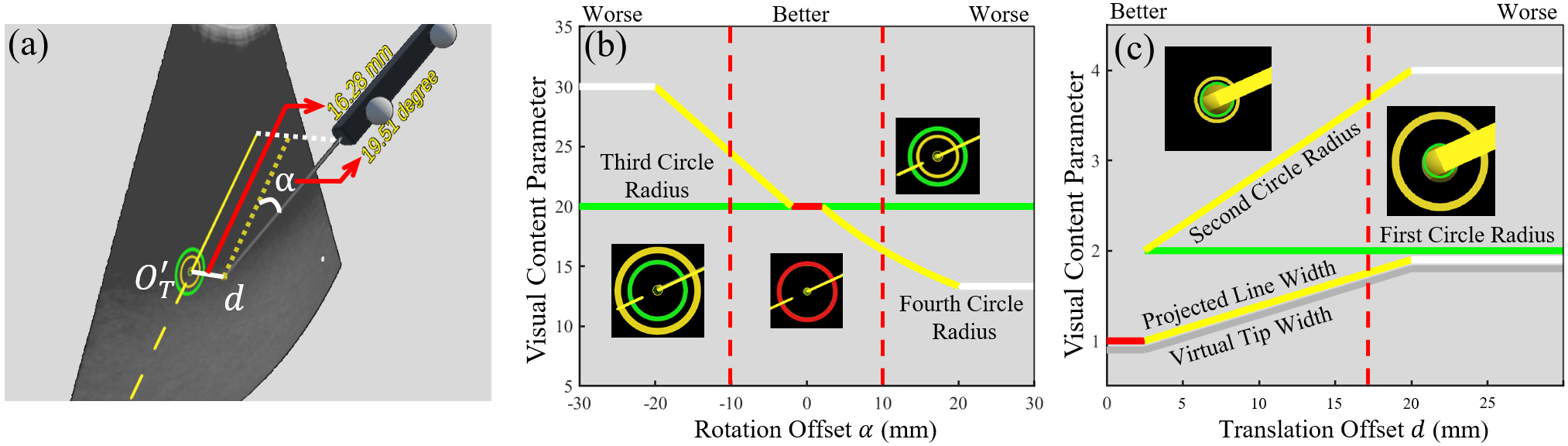}
    \caption{Visualization contents during in-plane punctuation. (a) Calculation of offset between biopsy needle and ultrasound image. Conversion from (b) rotation  and (c) translation offset values to visualization status. Here, the color of the line indicates the display color of the target visual content.}
    \label{fig:InplaneinsituNavigation}
\end{figure}

During in-plane puncture, the navigation information serves to help align the biopsy needle and the ultrasound probe while indicating the trajectory direction. (see Fig. \ref{fig:InplaneinsituNavigation}(d)).
Given the current ultrasound image with origin $O_I$ and normalized norm $n_I$, and the position $O_T$ and direction $d_T$ of the biopsy needle's tip, we project the needle to the image plane to create a virtual shadow.
The projected origin and direction are calculated as follows:

\begin{equation}
    \begin{cases}
        O'_T=O_I+((O_I-O_T)\cdot n_I) n_I\\
        d'_T=normalize(d_T-(d_T\cdot n_I)n_I)
    \end{cases}
\end{equation}

We represent the path already traversed as solid, while the future path is shown as dashed.
Two groups of concentric circles following projected needle tip $O'_T$ indicate the misalignment between the biopsy needle and the ultrasound image in terms of direction and rotation.
Smaller concentric circles are used to indicate offset. 
As the needle tip approaches the ultrasound image, the second circle ($R_2$) becomes progressively smaller and eventually aligns with the first circle ($R_1$). 
The rotational misalignment is depicted similarly using the third ($R_3$) and fourth ($R_4$) circles.
To ensure that the virtual objects do not obstruct the ultrasound image, the virtual tooltip and projected trajectory become thinner as the tooltip approaches the ultrasound image.
Detailed relationship between the parameters for the displayed contents and the image-needle misalignment is shown in Fig. \ref{fig:InplaneinsituNavigation}(b, c).

During out-of-plane puncture, we visualize the distance from the needle tip to the image plane and the potential hitting point of the biopsy needle on the ultrasound image.
Given the same hypothesis as in the previous part, this information can be calculated as:

\begin{equation}
    \begin{cases}
        d = (O_I-O_T)\cdot n_I\\
        P = O_T + d \cdot d_T
    \end{cases}
\end{equation}

\begin{figure}[htpb]
    \centering
        \includegraphics[width=0.9\columnwidth]{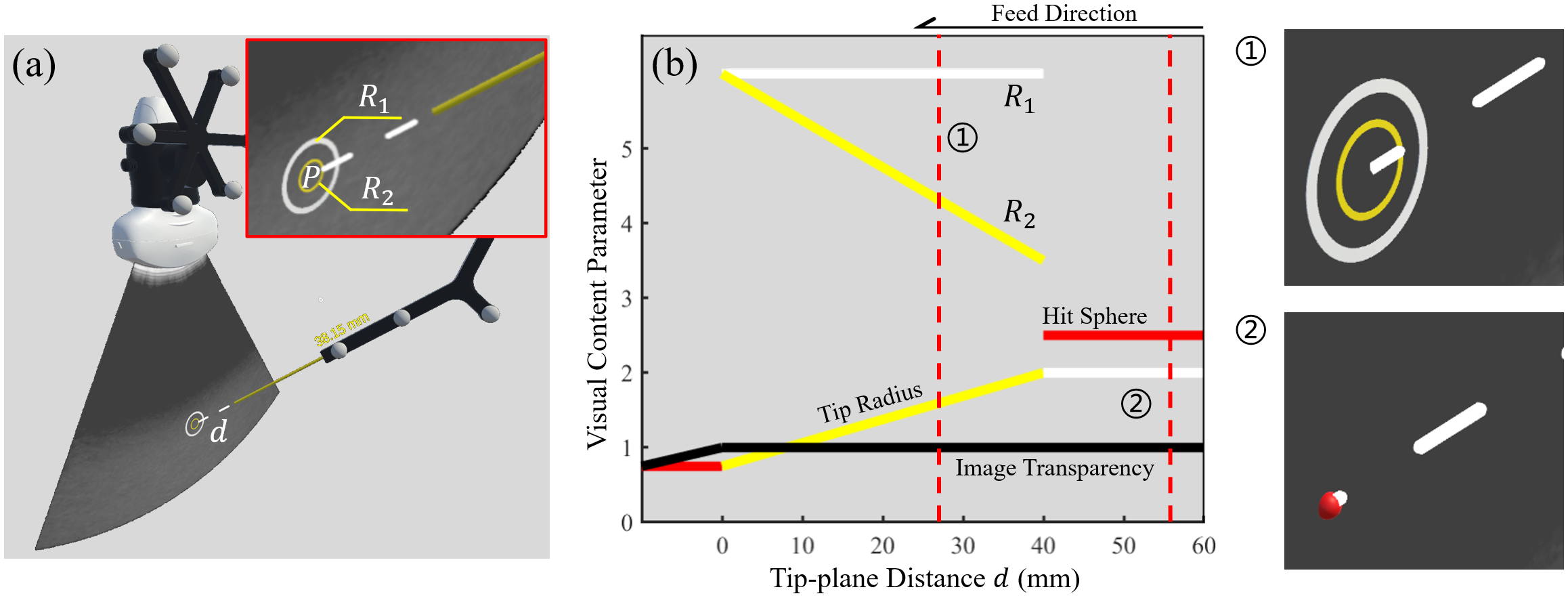}
    \caption{Visualize contents during out-of-plane punctuation. (a) Visual effect during out-of-plane navigation. (b) Conversion from the spatial relationship between the ultrasound probe and the biopsy needle to the status of visual contents.}
    \label{fig:ExplaneinsituNavigation}
\end{figure}

In this mode, when the needle tip is far from the ultrasound probe, a small sphere indicates the target hitting point of the biopsy needle on the image plane.
As the needle tip approaches the image plane, two concentric circles replace the sphere to show the tip's distance to the image plane.
The closer the needle tip is to the image plane, the more the two circles overlap.
The radius of the virtual needle tip becomes smaller in this procedure to ensure the observation of the alignment of the virtual needle and the needle's spot in the ultrasound image after hitting.
When the needle tip touches the image plane, two circles overlap and disappear, the virtual tooltip turns red, and the ultrasound image becomes transparent to visualize the extruded needle tip.
Fig. \ref{fig:ExplaneinsituNavigation} shows the relationship between the distance $d$ and the visual content status during out-of-plane punctuation.

The effect of the proposed visualization method is demonstrated in Fig.\ref{fig:VirtualAlignedRealScene}.
Virtual information is overlayed on the captured RGB frames of the real scene.

\section{Results} \label{sec3}
\subsection{Latency Analysis}
\begin{figure}[htpb]
    \centering
        \includegraphics[width=0.9\columnwidth]{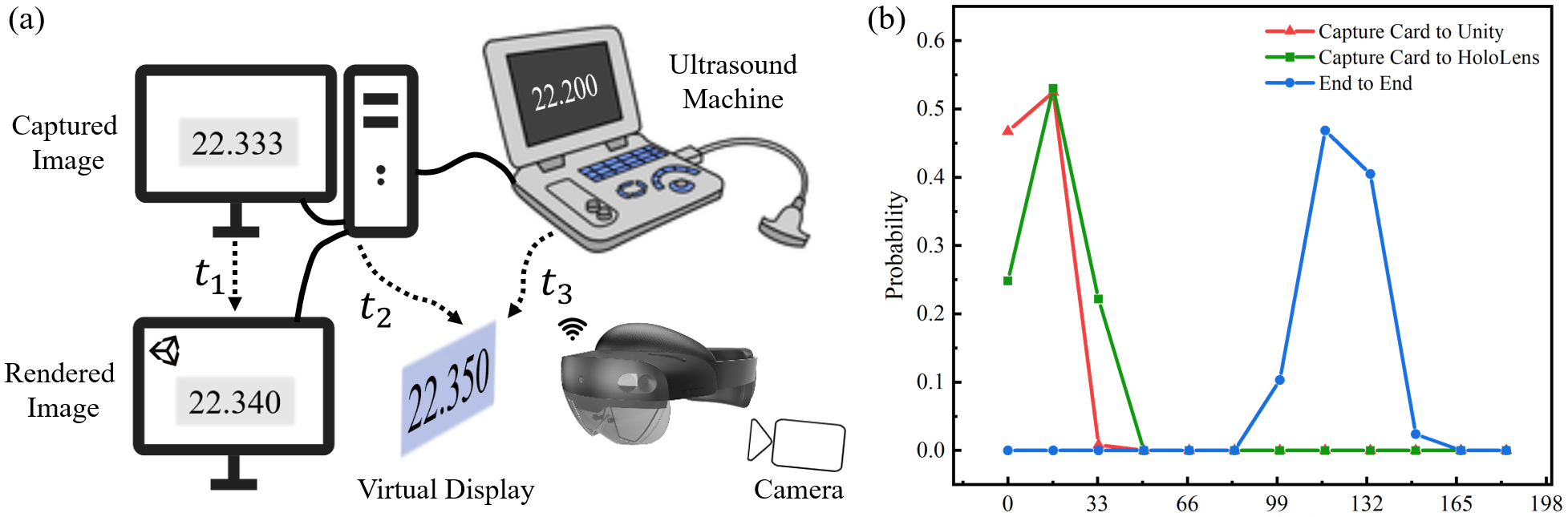}
    \caption{Evaluation of latency of different steps in the proposed system. (a) A camera at 60fps is used to record video from HoloLens 2 glass for latency analysis. The number displayed in the figure is in the unit of second. (b) Latency of the proposed system at different stages.}
    \label{fig:Latency}
\end{figure}

The latency of our proposed system was evaluated within a simulated ultrasound environment.
As the latency highly depends on the specific ultrasound and image capture card adopted, despite end-to-end latency from the ultrasound screen to the AR headset, we also considered the latency from the acquisition of ultrasound image to display in Unity and on HoloLens 2 (see Fig. \ref{fig:Latency}).
We conducted the latency analysis using a stopwatch displayed on the first screen to simulate the ultrasound machine.
The stopwatch data were captured by the image capture card in the computer running our proposed system.
The acquired ultrasound image was displayed on the screen immediately upon capture, then it passed through the entire system and was displayed in Unity and on the AR headset. 
We recored video using a camera behind HoloLens 2 glass at 60 frames per second (fps) for 10 seconds to analyze system latency.
In this analysis, we used ultrasound images of the largest size, which were $1053\times 604$ pixels, along with the mask in the same size, resulting in a data size of 1.21 MB per frame.

As a result, there was $t_1 = 9.02\pm 8.59 ms$ and $t_2 = 16.22\pm 11.45 ms$ latency from the access of the screen image to display in Unity and AR headset, respectively.
The end-to-end latency from the first screen display to the AR headset was $t_3 = 122.49\pm11.61ms$.
This analysis demonstrates that our system can visualize ultrasound images in AR with an additional latency of only $16.22\pm 11.45$ ms, making it suitable for real-time applications.

\subsection{Navigation Accuracy}
\begin{figure}[htpb]
    \centering
        \includegraphics[width=0.55\columnwidth]{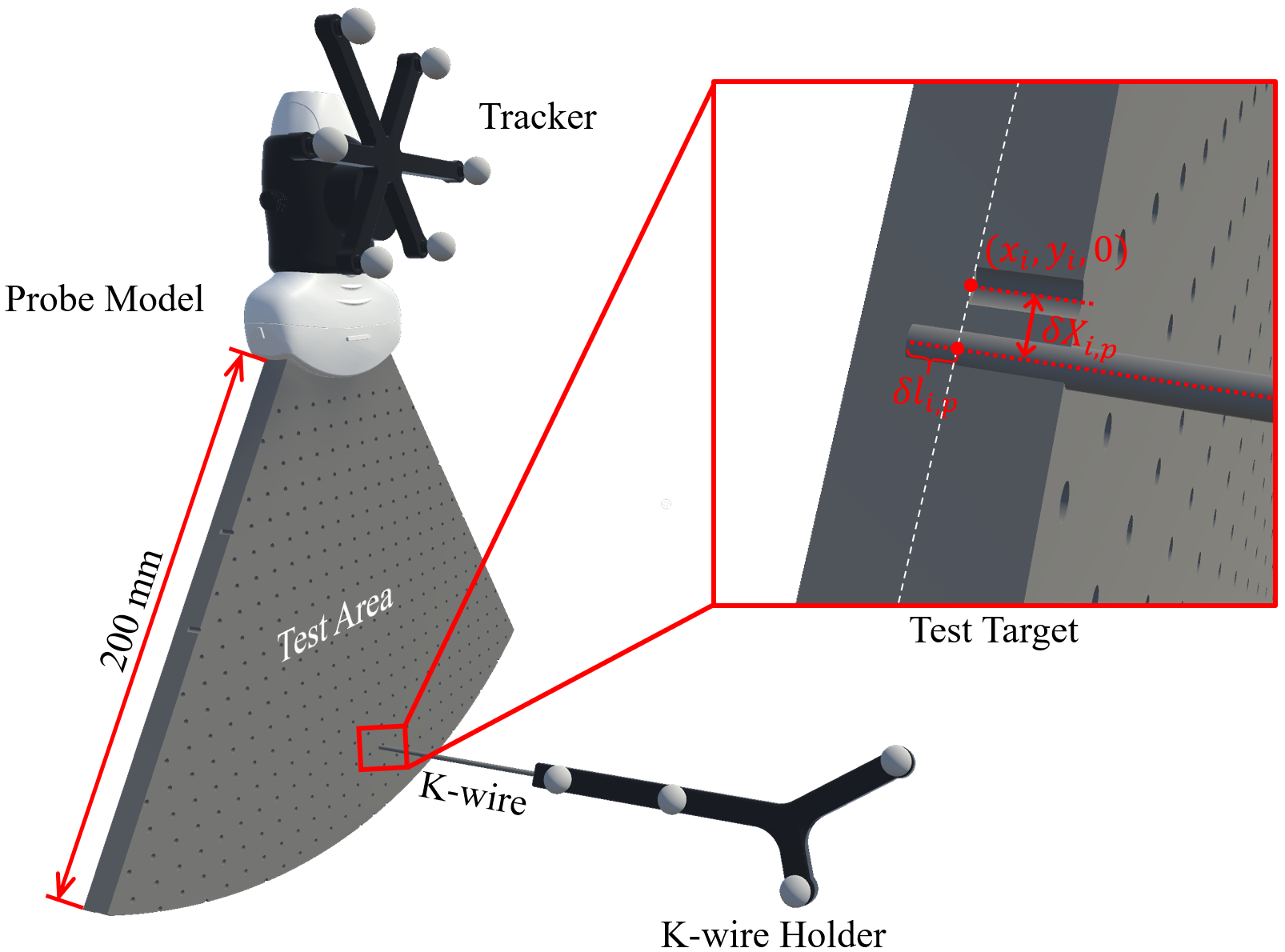}
    \caption{Custom structure to test biopsy navigation accuracy and stabilty over the imaging area witin $200mm$ depth.}
    \label{fig:AccuracyTestingStructure}
\end{figure}

We evaluated the accuracy of biopsy navigation using a custom structure, as illustrated in Fig. \ref{fig:AccuracyTestingStructure}.
As the tracking accuracy of a single infrared tool has been illustrated in previous works \cite{martin2023sttar, von2022ultrarsound}, we focus mainly on the accuracy and stability of visual navigation contents, with relevant to the ultrasound image.
The structure included a 3D-printed ultrasound probe and a testing area resembling the valid image area at a depth of $200mm$. 
Holes with positions $(x_i, y_i, 0)$ were created in the testing area, with half of each hole remaining solid.
A k-wire was used to simulate the biopsy needle, and its tip was positioned $120mm$ from the center of the infrared tool.
During the test, the k-wire was inserted into the hole, with the tip touching the solid surface.
By tracking the infrared tool on the ultrasound probe and the biopsy needle, we estimated the poses of the ultrasound image ($T_{image}$), the origin of the k-wire ($O_{kwire}$), and the direction of the k-wire ($d_{kwire}$). 
We calculated the intersection point of the k-wire and the ultrasound image by solving the following equation.

\begin{equation}
    T_{image,i,p}
    \begin{bmatrix}
        x_{measure,i,p} \\
        y_{measure,i,p} \\
        0
    \end{bmatrix}
    =O_{kwire,i,p}+(l+\delta l_{i,p})d_{kwire,i,p}
\end{equation}

Here $p$ refers to the $p$th sample for each target.
We evaluated navigation accuracy using two-dimensional navigation information accuracy $\delta X_{i,p} = ||(x_{measure,i,p},y_{measure,i,p})-(x_i,y_i)||_2$ and depth accuracy along the feeding direction $\delta l_{i,p}$.
In total, 353 target positions were included in this experiment, with adjacent testing positions spaced at $10mm$. 
For each target, we collected 200 continuous frames of data for analysis.

\begin{figure}[t]
    \centering
        \includegraphics[width=1.0\columnwidth]{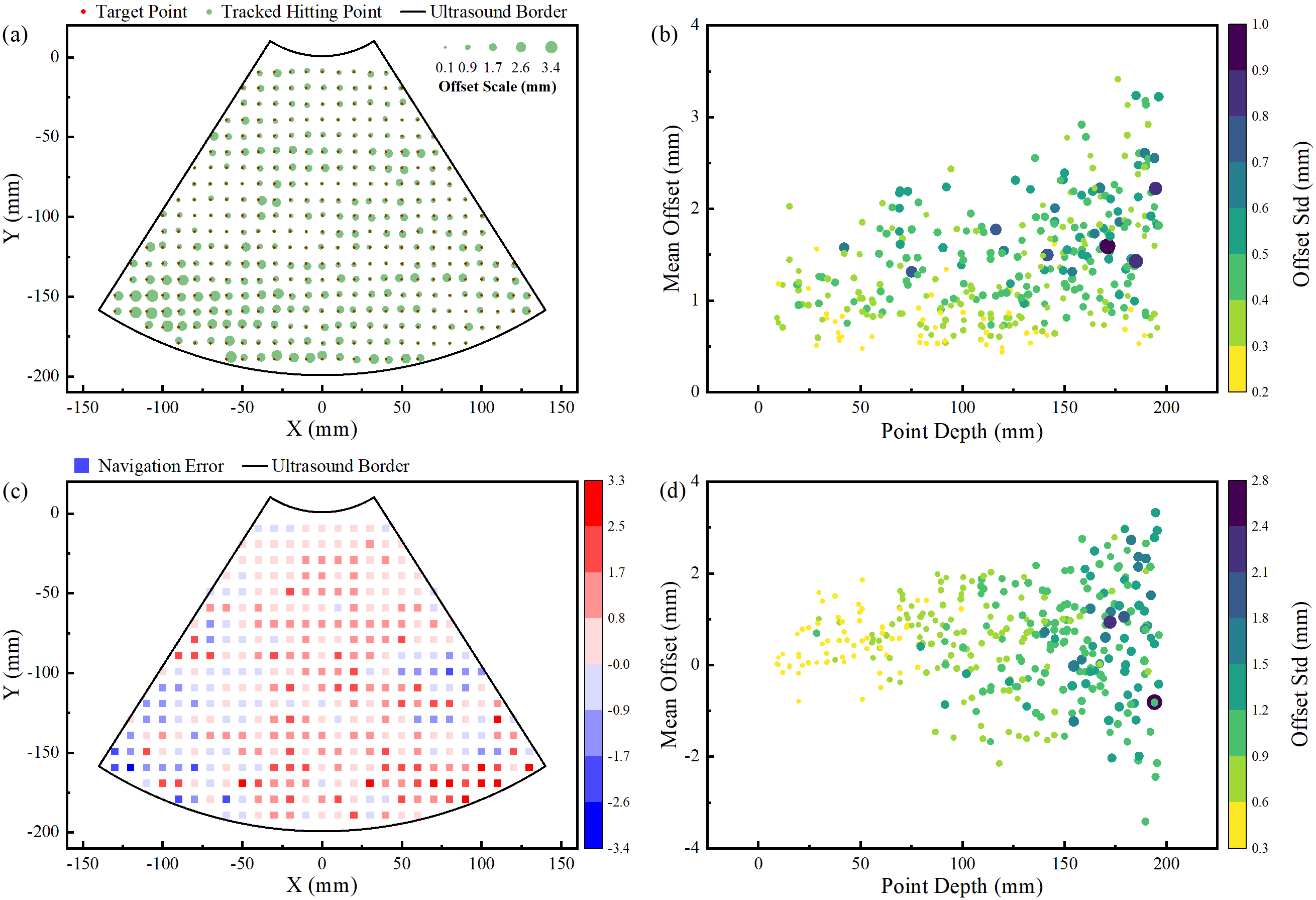}
    \caption{Accuracy of visualize contents in and out of the image plane. (a,b) Navigation accuracy in the image plane. (c,d) Navigation accuracy out of the image plane. Size and color of points in (b) and (d) indicated the standard deviation of the error over 200 continuous frames.}
    \label{fig:TrackingAccuracy}
\end{figure}

The distribution of the tracking error in the image plane, and the relationship between the tracking error and the target depth are shown in Fig. \ref{fig:TrackingAccuracy}.
Fig. \ref{fig:TrackingAccuracy} (a) shows the mean position of the detected intersection point in all detected frames $P_i = Mean_{i\in (1,200)}((x_{measure,i,p},y_{measure,i,p}))$ at different test targets. 
To better show the offset scale, the point size in the figure is set linearly to the offset distance $d = \|P_i-(x_i,y_i)\|$.
On average, we observed an offset of $1.23\pm 0.68mm$ at all test points.
Since most biopsies occur within $50mm$ depth, we further divide the depth range by $50mm$ intervals.
In general, we found offsets of $0.90\pm 0.36$, $1.10\pm 0.58$, $1.01\pm0.55$, and $1.60\pm 0.75$ millimeters for target depths in the ranges $[0,50)$, $[50,100)$, $[100,150)$, and $[150,200)$ millimeters, respectively.
Fig.\ref{fig:TrackingAccuracy} (b) presents the distribution of the absolute offset $X_{i,p}$ with regard to target depth in the ultrasound image.
In this figure, each point represents the mean absolute error (MAE) at one test target, with the offset standard deviation indicated by the size and color of the point.
On average, $1.37mm$ MAE was found, which was $1.04mm$ within the $50mm$ range.
An increase in maximum mean offset as the target depth increases was observed, while the minimum one remained similar, which can be explained by the discrete distribution of the sensor depth value \cite{ungureanu2020hololens}.
In terms of MAE out of the plane, $0.95\pm0.70mm$ was presented for all depths, with $0.52\pm0.45$, $0.80\pm0.51$, $0.92\pm0.57$, and $1.20\pm 0.84$ millimeter, respectively, for increasing depth ranges with $50mm$ interv.
Both the mean and deviation of the tracking error increased with target depth.
Moreover, we observed a distortion on both sides of the image plane in the case of in-plane offsets (see Fig. \ref{fig:TrackingAccuracy} (a)), with $78.7\%$ of measured points being closer to the center of the image on the x axis.
This phenomenon is consistent with the results of previous studies on infrared tool tracking \cite{martin2023sttar}.

\subsection{Use Case}
\begin{figure}[t]
    \centering
        \includegraphics[width=1.0\columnwidth]{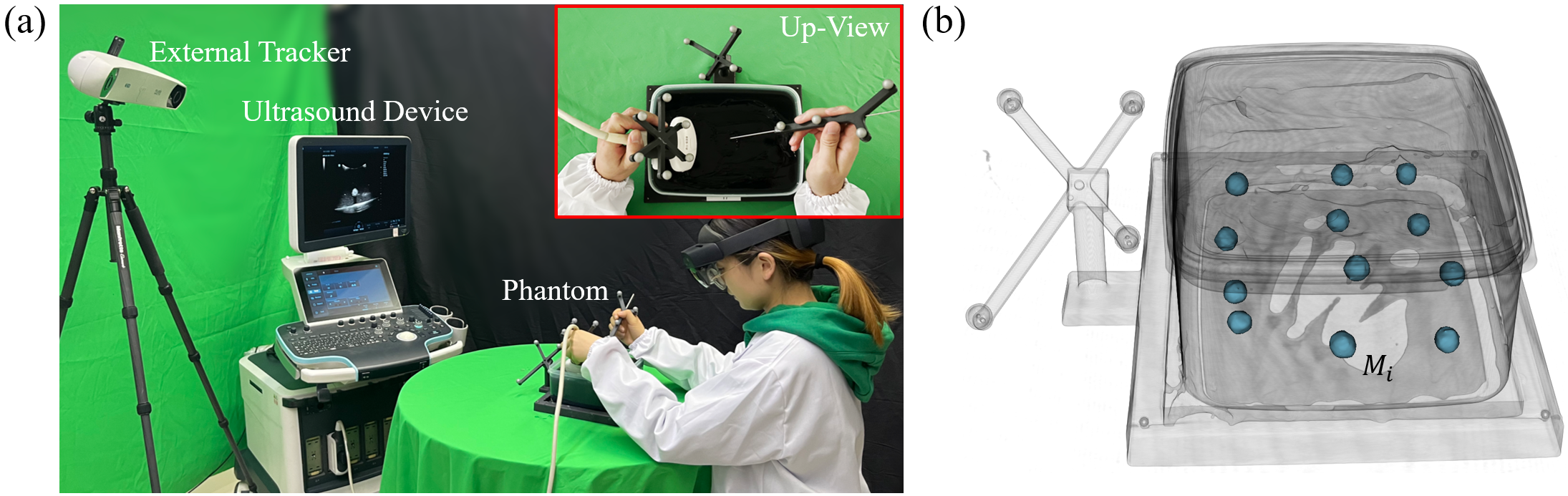}
    \caption{Setup of the use case study. (a) Environment setup to evaluate biopsy accuracy using AR navigation. An external tracking device is used to calculate the punctuation error once a biopsy procedure is finished. (b) CT scan of the used phantom to simulate tumor simulation. Targets with $15mm$ diameter is made with tapioca flour to simulate tumors.}
    \label{fig:UseCaseRealScene}
\end{figure}

To further illustrate the impact of our proposed system and different visualization methods on biopsy navigation, we conducted the following use case study (see Fig. \ref{fig:UseCaseRealScene}).
We fabricated a pahntom using gelation with $7\%$ concentration and created biopsy targets with a diameter of 15mm using tapioca flour.
These targets were suspended at a depth of approximately 40mm within the phantom.
The phantom was securely mounted on a custom structure equipped with four retro-reflective markers for localization.
We performed CT scans at 0.625mm spatial resolution to determine the central positions of the targets within the tracking space (denoted as $M_i$).
During the experiment, participants were tasked with puncturing the k-wire into the center of the targets under the guidance of the AR system.
To track both the phantom and the k-wire for performance analysis, we employed an external tracking system (NDI Polaris Spectra).
Once a punctuation is finished, the origin $O_T$ and the normalized direction $d_T$ of k wire are calculated in the phantom space.
The punctuation is then evaluated with the directional and depth offsets:

\begin{equation}
    \begin{cases}
\delta_{direction} = \|d_T \times (M_i - O_T)\|\\
\delta_{depth} = (O_T+l\cdot d_T-M_i)\cdot d_T
    \end{cases}
\end{equation}

Here, $l$ represents the length of the k-wire.
Directional error indicates how accurate the participant points the biopsy needle at the target, while depth error indicates whether the puntuation is ended properly after hitting the target center.
As a biopsy needles have long biopsy areas, direction area is more important to determine whether the biopsy is success or not.
Given a target with radius $r$, success of the biopsy can be determined by the comparsion $\delta_{depth} < r$.

\begin{figure}[htpb]
    \centering
        \includegraphics[width=1.0\columnwidth]{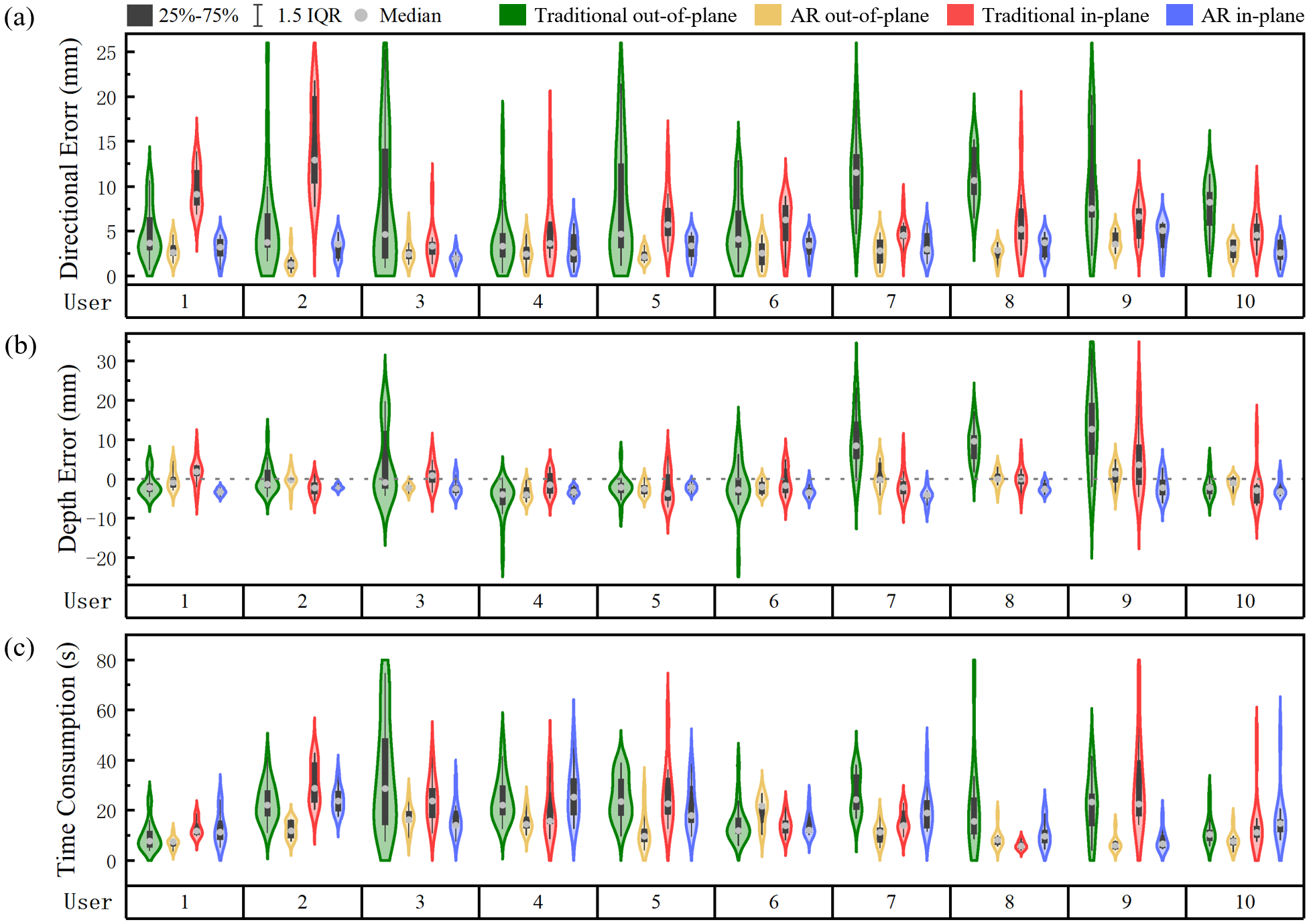}
    \caption{Results of the use case study. (a) Directional error, (b) depth error and (c) time consumption in a use case study including 10 participants. Each participant performed 10 biopsy simulations with and without AR. Performance under both in-plane and out-of-plane biopsy was evaluated.}
    \label{fig:UseCaseResult}
\end{figure}

We enlisted ten participants, all of whom were newcomers to ultrasound-guided surgery (7 males, 3 females, $M_{age}=24.80$,$SD_{age}=3.15$), to demonstrate how our system can assist novice surgeons in performing biopsies. 
We considered four different visualization methods, including in-plane and out-of-plane navigation, with and without AR guidance.
For each participant, the sequence of the four visualization methods was randomized in previous.
Five practice punctures were performed in each scenario to familiarize the participants with the system. 
During the formal experiment, ten punctures were completed in each mode. After each puncture, we recorded the directional and depth errors ($\delta_{direction}$ and $\delta_{depth}$) to calculate the errors, as well as the time taken ($t$).
Participants were not allowed to completely remove the k-wire from the model once the puncture had begun to prevent repeated operations.

The results of the use case experiment are shown in Fig. \ref{fig:UseCaseResult}.
As the error distributions did not follow a normal distribution, we described the performance using median values and interquartile ranges (IQR).
Overall, for out-of-plane punctures, we observed a directional error of $9.02mm$ with an IQR of $7.64mm$ across all trials, which was reduced to a median of $2.58mm$ with an IQR of $1.60mm$ when our proposed system was used. 
In the case of in-plane punctures, the median error without AR was $5.76mm$ with an IQR of $4.48mm$, and with AR, it was $3.04mm$ with an IQR of $2.30mm$.
Regarding depth error, for out-of-plane punctures, we found a median error of $4.49mm$ with an IQR of $7.66mm$, which was improved to a median error of $1.85mm$ with an IQR of $2.67mm$ with the use of our system. 
For in-plane punctures, a median error of $2.53mm$ with $3.07mm$ was found without AR, which was $3.01mm$ and $1.71mm$ with AR navigation.

In terms of time, traditional out-plane punctures took 20.00 seconds (IQR = 20.02) on average, which was reduced to 10.06 seconds (IQR = 8.55) with our system.
Similar performance was found for in-plane operation with ($Median=15.49$, $IQR=10.35$ seconds) and without ($Median=15.98$, $IQR=12.35$ seconds) AR.

If we consider a tumor with a radius of $5mm$, the biopsy success rate was $26\%$ without AR for out-of-plane punctures and $45\%$ for in-plane punctures. 
In contrast, with the use of AR, the success rates increased significantly to $98\%$ for out-plane punctures and $93\%$ for in-plane punctures. 
In our experiment, where targets with a radius of 15mm were used, we achieved a $100\%$ success rate in both scenarios.

\section{Discussion} \label{sec4}
Ultrasound-guided biopsy procedures are inherently limited by the lack of spatial information outside the image plane.
In this study, we introduced a novel augmented reality (AR) system designed to provide in-situ real-time ultrasound-assisted biopsy navigation.
Our system architecture involves the acquisition of ultrasound images via an image capture card, processing and streaming of these images to Unity for rendering, and then forwarding the rendered result to the HoloLens 2 headset for visualization through holographic remoting. 
This approach significantly reduces display latency, due to fast network communication within a single high-performance computer (HPC), rapid rendering speed, and minimized redundant data when synchronizing information from the HPC to the AR headset.
Additionally, we implemented infrared tool tracking on the AR headset during remote rendering, minizing time consumption for sensor data acquisition and tracking result synchronization.
To improve biopsy navigation, we provided distinct visual cues for in-plane and out-of-plane punctures, including offset cues and needle direction for in-plane punctuation, needle-image distances and potential intersection points for out-of-plane punctuation.

First, we evaluated the latency of our proposed system.
Only $16.22\pm 11.45$ milliseconds was required to transmit an image acquired by the HPC to the AR headset in the worst case.
In Costa's work, which employed on-device resources for ultrasound navigation \cite{costa2023augmented}, $23.681\pm 0.637 ms$ was used for ultrasound data reception and $14.247\pm 0.653ms$ for image rendering.
Haxthausen et al. reported a latency of $16 ms$ between the ultrasound image display on the workstation and the AR headset \cite{von2022ultrarsound}, which occasionally falls into latency as high as $40ms$.
The low latency that Haxthausen's work achieved is due to the small size of their ultrasound image ($512\times 512\ pixels$), and no need for mask information due to the linear probe.
Although $4.85$ times more data for image and mask information are streamed in our system, only $16.22ms$ extra latency is introduced.
The low latency our system achieved even under large ultrasound images is due to the remote rendering adopted.
The synchronization and rendering is kept within an HPC, greatly improving the speed.
After rendering, most of the redundant information is compressed, which further decreases the bandwith needed to synchronize data from HPC to AR headset.

As the majority of navigation information is projected onto the image plane in our system, we assessed navigation accuracy within and outside the image plane.
For points on the image plane within $200$mm, we observed a mean offset of $1.23\pm 0.68$mm between the real and tracked intersection positions.
When the needle tip touched the plane, it was detected to be $0.95\pm 0.70$mm out of the plane.
Both in-plane and out-of-plane errors increased with depth, although their relationships with depth offset were different (see Fig.\ref{fig:TrackingAccuracy} (b,d)).
The standard deviation of the out-of-plane offset basically increased with depth; however, even at a depth of $200$mm, small deviations still existed for some targets.
The primary source of tracking error for the infrared tool is the limited accuracy of the depth sensor.
Since the infrared spheres are arranged parallel to the image plane, most tracking instability is reflected in the depth direction.
This is noteworthy because the biopsy needle is nearly perpendicular to the image plane during out-of-plane punctures, allowing for high in-plane stability even when the target is deep.
Taking into account the volumetric effect of ultrasound and the $3-6$mm thickness of ultrasound images \cite{zhong2008image}, our measured accuracy ensures the alignment of virtual trajectories and their effect of real trajectories in ultrasound images.
This accordance may help surgeon align the virtual information with their experience.

In our use case study, we recruited 10 participants with no prior experience in ultrasound-guided biopsies to perform biopsy simulations under both in-plane and out-plane circumstances.
In general, our proposed system significantly improved the biopsy success rate, increasing it from $26\%$ to $98\%$ for out-of-plane punctures and from $45\%$ to $93\%$ for in-plane punctures, assuming hypothetical biopsy targets with a $5mm$ radius.
During data analysis, we segregated biopsy errors into directional and depth errors.
The former indicates the accuracy of the biopsy needle's alignment with the target, while the latter reflects where participants halted the biopsy needle during the experiment.
In out-of-plane punctuation, both the median and IQR of directional and depth errors improved for all participants.
Comparatively, in in-plane puncutures, all participants exhibited improvement in directional error median and depth error IQR, while seven participants demonstrated improvement in directional error IQR, and four participants showed improvement in depth error median.

Participants reported difficulties in locating the needle tip and determining its direction during out-of-plane biopsies when AR was not used.
These challenges explain the highest directional and depth error among all four modes.
With the adoption of AR, visual cues were provided to indicate biopsy direction and offer guidance when the needle intersected with the image plane.
Consequently, performance improved in terms of both directional and depth error, and a reduction in procedure time by $49.7\%$ was achieved.
In in-plane operations, ultrasound images inherently provided spatial relationships between the biopsy needle and target tissue in both direction and depth, resulting in improved performance. 
However, in this mode, depth information was already well presented in the ultrasound image.
Additional virtual guidance helped participants align the biopsy needle with the image plane, leading to better directional performance, but only marginal improvements in depth error and procedure time.
In summary, the use case study demonstrates the system's ability to assist in aligning the biopsy needle with the ultrasound image in in-plane punctures and in locating the needle in out-of-plane punctures.

Furthermore, the proposed system offers more than just improved biopsy performance with in-situ ultrasound information.
The incorporation of remote rendering into the framework facilitates the seamless integration of ultrasound image processing methods. 
This opens the door to providing visual cues such as recognized targets, segments, and reconstructed three-dimensional images to aid surgeons in comprehending ultrasound images.

\section{Conclusion} \label{sec5}
This paper introduces a framework for in-situ ultrasound biopsy navigation characterized by low latency and high accuracy.
It leverages holographic remoting and infrared tool tracking to enable multiple tool tracking capabilities while streaming ultrasound images to augmented reality environments with minimal delay.
The framework employs distinct visualization methods tailored for in-plane and out-of-plane biopsy procedures.
In practice, ultrasound images acquired by a computer can be seamlessly streamed to the AR environment in just $16.22\pm 11.45$ milliseconds.
The system achieves precise in-situ navigation with an in-plane error of $1.23\pm 0.68mm$ and an out-plane error of $0.95\pm 0.70mm$ within a $200$ millimeter depth range.
In a use case study involving 10 participants, the proposed system significantly enhances biopsy success rates, achieving a 2.07-fold improvement for in-plane operations and a 3.77-fold improvement for out-of-plane operations. 

\bmhead{Acknowledgments}
This research was supported by NSFC(U20A20389), National Key R\&D Program of China (2022YFC2405304) and Medivis. Inc.

\bibliography{sn-bibliography}
\end{document}